\documentclass[conf]{new-aiaa}
\usepackage[utf8]{inputenc}

\usepackage{graphicx}
\usepackage{amsmath}
\usepackage[version=4]{mhchem}
\usepackage{siunitx}
\usepackage{longtable,tabularx}
\usepackage[colorinlistoftodos,prependcaption,textsize=tiny]{todonotes}
\usepackage{algorithm}
\usepackage{algorithmic}
\usepackage{bm}
\usepackage{subfig}
\usepackage{rotfloat}
\usepackage{booktabs}
\usepackage{multirow}

\newcommand{\algorithmicinput}{\textbf{input}}
\newcommand{\algorithmicoutput}{\textbf{output}}
\newcommand{\INPUT}{\item[\algorithmicinput]: }
\newcommand{\OUTPUT}{\item[\algorithmicoutput]: }

\title{An efficient application of Bayesian optimization to an industrial MDO framework for aircraft design}

\author{R\'emy Priem\footnote{Ph.D.\ Candidate, Information Processing and Systems Department \& Complexes Systems Engineering Department, remy.priem@onera.fr, Student AIAA Member}}
\affil{ONERA, DTIS, Université de Toulouse, Toulouse, France}
\affil{ISAE-SUPAERO, Université de Toulouse, Toulouse, 31055 Cedex 4, France}
\author{Hugo Gagnon,\footnote{Engineering Professional, Advanced Design, hugo.gagnon@aero.bombardier.com.} Ian Chittick,\footnote{Engineering Specialist, Advanced Aerodynamics, ian.chittick@aero.bombardier.com.} and St\'ephane Dufresne\footnote{Senior Engineering Specialist, Specialized Aircraft, stephane.dufresne@aero.bombardier.com.}}
\affil{Bombardier Aviation, Dorval, Qu\'{e}bec, H9P 1A2, Canada}
\author{Youssef Diouane\footnote{Associate Professor, Complexes Systems Engineering Department, youssef.diouane@isae-supaero.fr.}}
\affil{ISAE-SUPAERO, Université de Toulouse, Toulouse, 31055 Cedex 4, France}
\author{Nathalie Bartoli\footnote{Senior researcher, Information Processing and Systems Department, nathalie.bartoli@onera.fr, AIAA Member.}}
\affil{ONERA, DTIS, Université de Toulouse, Toulouse, France}

\begin{document}

\maketitle

\begin{abstract}
The \textit{multi-level}, \textit{multi-disciplinary} and \textit{multi-fidelity} optimization framework developed at Bombardier Aviation has shown great results to explore efficient and competitive aircraft configurations. 
This optimization framework has been developed within the Isight software, the latter offers a set of ready-to-use optimizers. 
Unfortunately, the computational effort required by the Isight optimizers can be prohibitive with respect to the requirements of an industrial context.
In this paper,  a \textit{constrained Bayesian optimization} optimizer, namely the \textit{super efficient global optimization with mixture of experts}, is used to reduce the optimization computational effort. The obtained results showed significant improvements compared to two of the popular Isight optimizers. 
The capabilities of the tested \textit{constrained Bayesian optimization} solver are demonstrated on \textit{Bombardier research aircraft configuration} study cases.
\end{abstract}

\section{Nomenclature}

{\renewcommand\arraystretch{1.0}
\noindent\begin{longtable*}{@{}l @{\quad=\quad} l@{}}
$d$ & the number of design variables \\
$m$ & the number of inequality constraints \\
$\Omega$ & the domain of the design variables, i.e.,  $\Omega \subset \mathbb{R}^d$\\
$f$  & the objective function, i.e., $f: \mathbb{R}^d \mapsto \mathbb{R}$ \\
$\bm{g}$ & the constraints function, i.e.,  $\bm{g}: \mathbb{R}^d \mapsto \mathbb{R}^m$\\
$\mu_s^{(l)}$ & the prediction of the Gaussian process of a given function $s: \mathbb{R}^d \mapsto \mathbb{R}$ build with $l$ samples\\
$\sigma_s^{(l)}$ & the uncertainty of the Gaussian process of a given function  $s: \mathbb{R}^d \mapsto \mathbb{R}$ build with $l$ samples \\
$\alpha_f^{(l)}$ & the acquisition function of the objective function $f: \mathbb{R}^d \mapsto \mathbb{R}$\\
$\alpha_{\bm{g}}^{(l)}$ & the feasibility criterion of the constraint functions $\bm{g}: \mathbb{R}^d \mapsto \mathbb{R}^m$\\
$\Omega^{(l)}_{\bm{g}}$ & the approximated feasible domain defined by the feasibility criterion $\alpha_{\bm{g}}^{(l)}: \mathbb{R}^d \mapsto \mathbb{R}^m$ \\
$C_{L_{max}}$ & the low speed maximum lift coefficient
\end{longtable*}}

\section{Introduction}
\lettrine{B}{ombardier} aviation \textit{multi-disciplinary optimization} (MDO) framework has been continuously evolving over the past decade to ensure that future development programs result in highly efficient and competitive aircraft~\cite{Piperni2004}.
In order to achieve this vision, Bombardier has developed a multi-level, multi-fidelity MDO framework to mature an aircraft configuration from early design stage to detailed design~\cite{Piperni2013}.  

The first element of this framework is called \textit{Conceptual MDO} (CMDO).
The main objective of the CMDO framework is to enable the design team to explore the design space assuming a set of realistic constraints established from previous aircraft development programs and marketing requirements.
The main deliverable of this process is an aircraft configuration, which includes the initial sizing of the major structural components and systems.
This conceptual aircraft configuration then generates a set of design requirements that will be validated using high fidelity frameworks, namely the \textit{Preliminary MDO} (PMDO) and the \textit{Detailed MDO} (DMDO).
The second element corresponds to the PMDO framework analyzing the aerodynamic and structural characteristics of the wing.
Its main objectives are to refine the wing sizing and to validate the wing related conceptual requirements, such as drag polar and wing weight.
The two deliverables of this process are (1) a detailed wing geometry, external and internal, and (2) an initial definition of the wing structure. 

The context of this research is about the design process of a new aircraft configuration in an industrial setting.
The main inputs to initiate the design of a new aircraft platform, as well as inputs to the MDO framework, are a set of requirements from the marketing team (e.g.\ cruise mach number, design range, field performance and critical airport operations).
For each of these sets of requirements, or constraints to the MDO framework, there needs to be a convergence between the CMDO and PMDO framework to ensure a robust initial design point.
Since there could be multiple sets of requirements to be analyzed, the design team becomes rapidly limited by the time and number of CPU available.

The CMDO framework is built on a large set of low fidelity modules representing the aircraft disciplines.
All these modules include their own set of design variables, which increases the number of dimensions of the optimization problem.
Consequently, it makes the thorough exploration of the design space expensive in terms of function calls and CPU time, in the order of days.
Similarly, the PMDO framework also includes a large number of design variables required to define the wing geometry and the aero-structural models are time consuming to converge, in the order of weeks.
The motivation of this research is then to apply Bayesian Optimization (BO) methods ~\cite{MockusBayesianmethodsseeking1975} to reduce the number of functions calls  and accelerate the convergence time for both MDO frameworks.
To do so, the SEGOMOE python tool-box is investigate against historical optimizers used at Bombardier aviation.
The main objective of this paper is thus to demonstrate that all the optimizers of the tool-box has good convergence properties and performs well on an industrial test case.  

This paper is organized into four sections: (1) detailed description of the models used in the CMDO  and PMDO frameworks, (2) introduction to the Bombardier Research Aircraft Configuration (BRAC) case study, (3) definition of the optimizer selected for the analysis of the case study and finally (4) the analysis of the results with the emphasis on the convergence in terms of CPU time and number of evaluations.

\section{An industrial multi-level, multi-fidelity and multi-disciplinary optimization process}

The MDO framework developed at Bombardier Aviation consists of three levels, corresponding to the three traditional stages of aircraft design: conceptual, preliminary, and detailed.
Only a brief description of the first two levels is given here. Refer to~\cite{Piperni2013} for more information.

\subsection{Conceptual multi-disciplinary optimization framework}

\textit{Conceptual MDO} (CMDO) is the first level of the multi-level MDO framework.
It accounts for all major aircraft-level disciplines, such as aerodynamics, structures, systems, weight and balance, performance, stability and control, and economics.
Due to the high number of disciplines, the level of fidelity is low and varies from empirical to simple physics-based methods.
For example, in the aerodynamics module, an in-house method is used to predict the low-speed maximum lift coefficient $C_{L_{max}}$ while a vortex-lattice method coupled with a 2D computational fluid dynamics code is used to compute high-speed trim drag.
Similarly, other disciplines can combine mixed-fidelity analyses depending on the desired accuracy and turnaround times.
Example usages of CMDO are design space exploration, optimization of marketing requirements and objectives, validation of business cases, assessment of technology insertion, down selection of promising configurations, and definition of performance targets~\cite{Piperni2013}.

Typical design variables of CMDO include the wing area and planform as well as the engine scaling factor in the case of a fixed engine architecture.
It is also possible to vary the thickness-to-chord ratios of the wing and chordwise positions of the spars at various spanwise locations.
Typical constraints include performance parameters such as the \textit{balanced field length} (BFL), \textit{approach speed} ($V_\textsubscript{ref}$), \textit{initial cruise altitude} (ICA), range for different missions, and the ability to take-off and land at critical airports.
Systems and geometric considerations such as landing gear integration and wingtip chord can also be included.
Typical objectives include \textit{maximum take-off weight} (MTOW), cost, and climate impact, or any combination thereof.
For this study only MTOW is used since it combines the effects of cost and climate impact through fuel burn.

The CMDO framework is implemented as a workflow in Isight \cite{vanisight2010}. Isight is a process-integration, design-optimization software developed and distributed by Dassault Systemes.
It comes with its own set of optimizers.

The output of CMDO is an input to \textit{Preliminary MDO} (PMDO), discussed next.

\subsection{Preliminary multi-disciplinary optimization framework}
\label{ssec:PMDO}

The \textit{preliminary MDO} (PMDO) is the second level of the multi-level MDO framework.
Compared to CMDO, the scope of the design space is narrowed, while the fidelity of the underlying analysis tools is increased.
PMDO focuses primarily on the coupling and trade-off between aerodynamics and structures.
The objective is to obtain wing \textit{outer mold lines} (OML) that are aerodynamically efficient in both high-speed and low-speed flight, as well as to provide adequate performance at critical off-design conditions.
At the same time, the OML must be structurally viable, yield low weight, and satisfy space allocation requirements.

\citet{Piperni2013} previously described the PMDO process in detail.
Although many of the components remain the same, the framework has continued to evolve, and a complete description falls outside the scope of this paper.
In brief, the aircraft surfaces are updated in a parametric CATIA V5 model.
Bombardier’s in-house structured mesh generator, MBGRID~\cite{PiperniBoudreau2003}, and Navier-Stokes flow solver, FANSC~\cite{Mohamed2009}, are used to compute the flow solutions at several high-speed design and static load case conditions.
The external loads are transferred to the wing structural mesh generated by AWSOM~\cite{Deblois2010}, Bombardier’s in-house wing sizing tool, which minimizes the primary wing-box weight while respecting various margin of safety constraints.
The low-speed characteristics are assessed using the semi-empirical Valarezo criteria~\cite{Valarezo1994} in conjunction with VSAERO, a 3D panel code \cite{maskew1982program}.
Finally, the high-speed drag and structural weight are related through a fractional change equation, as described in \cite{Piperni2007}.

\begin{figure}[htb!]
    \centering
    \includegraphics[width=0.95\textwidth]{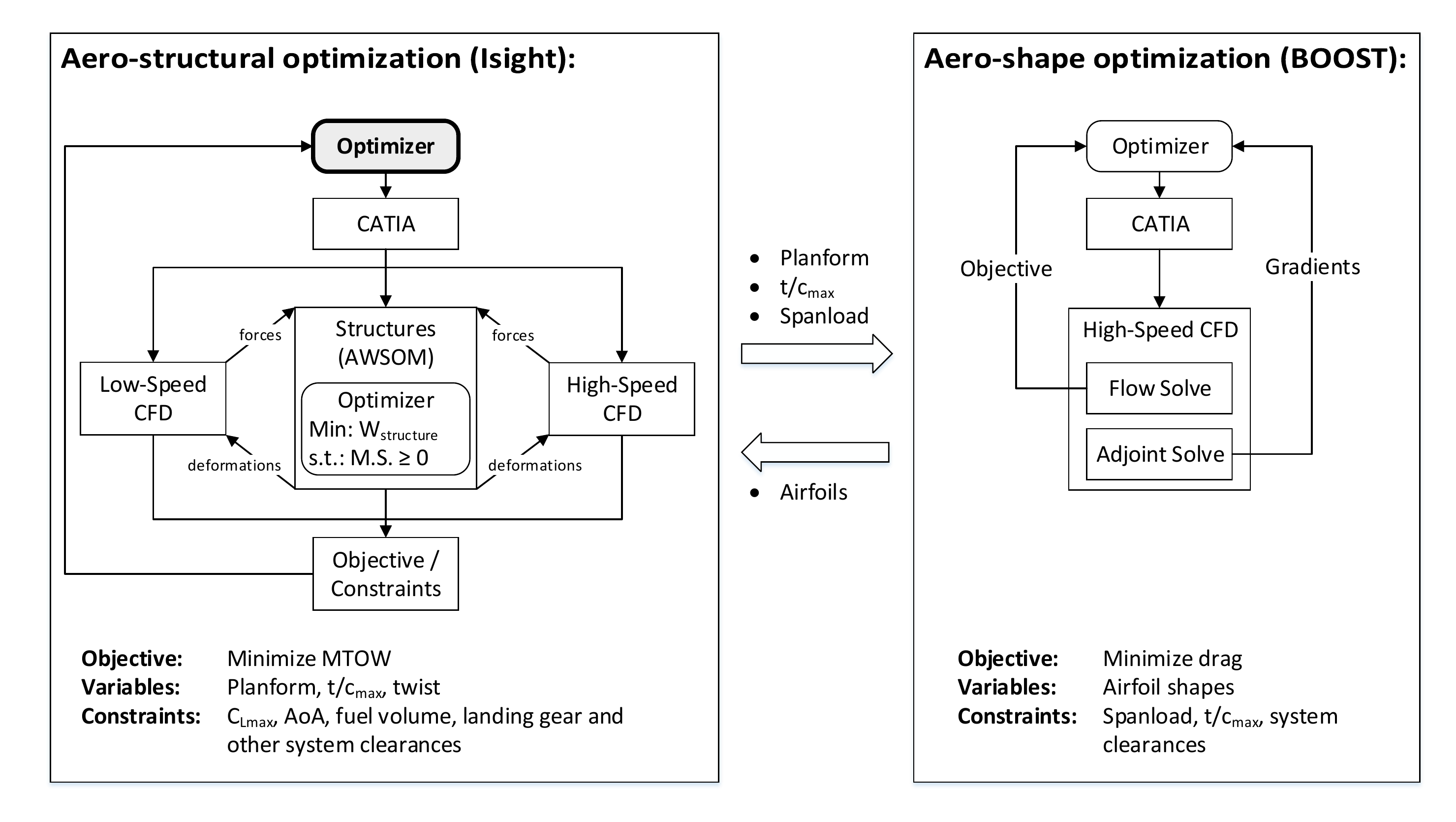}
    \caption{Simplified schematic of the hybrid-adjoint PMDO workflow.}
    \label{fig:PMDO_Framework}
\end{figure}

The architecture of the PMDO framework has changed significantly with the maturation of BOOST, an adjoint-based optimization framework used at Bombardier~\cite{Sermeus2010, Reist2019}.
The adjoint method is an efficient tool for computing the gradients needed for aerodynamic shape optimization, which involves long running simulation times and large numbers of design variables.
In order to integrate the aerodynamic adjoint in a multidisciplinary environment, a hybrid-adjoint MDO framework was developed~\cite{Deblois2013}, as depicted in Figure~\ref{fig:PMDO_Framework}.
In the first stage, an aero-structural optimization is performed in Isight by varying the wing plan-form, wing twist, and maximum thickness distribution (by scaling the wing profiles).
This is followed by an aero-shape optimization in BOOST, where the drag is minimized by varying the airfoil shapes at fixed thickness to maximum chord ratio and wing span-load.
This process is repeated in a sequential manner until convergence is achieved.
This paper will examine the top-level optimizer used to drive the aero-structural optimization, highlighted in bold in Figure~\ref{fig:PMDO_Framework}, which is the current bottleneck in the PMDO process.

\subsection{Bombardier research aircraft configuration}

\begin{figure}[htb!]
    \centering
    \includegraphics[width=0.5\textwidth]{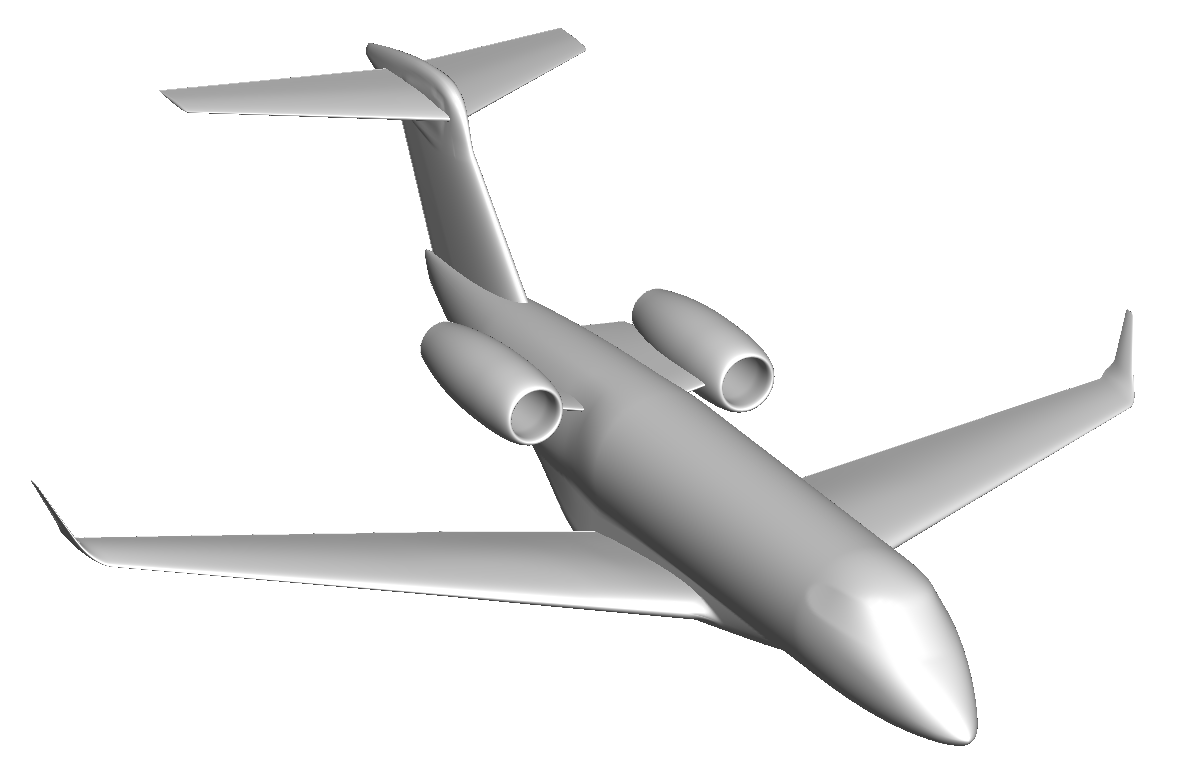}
    \caption{The baseline BRAC configuration.}
    \label{fig:BRAC_3DView}
\end{figure}

The test case considered for this work is representative of a small business aircraft, and is referred to as the \textit{Bombardier research aircraft configuration} (BRAC).
It is based on an early design of the Challenger 300 platform, and was developed to be shared with academic partners and for publication purposes.
An image of the baseline BRAC geometry is shown in Figure~\ref{fig:BRAC_3DView}.

\section{Bombardier research aircraft configuration optimization case}

In this section, the two level optimization process is introduced. First, we present the optimization of the BRAC model into the CMDO framework.
Then, the previously obtained result is used in the PMDO of the BRAC model.

\subsection{BRAC CMDO problem}

The BRAC CMDO problem is based on a modified version of the initial Challenger 300 marketing requirements and objectives.
It is thus representative of a typical industrial problem, yet here the design space is purposefully relaxed to test the global optimization capabilities of each optimizer.
For confidentiality reasons the exact bounds on the design variables and constraints are not given here; instead, in the following sections their value will be normalized between $0$ and $1$.

Formally, the CMDO problem is:
\begin{equation}
    \min\limits_{\bm{x}\in [0,1]^d}\left\{ \text{MTOW}(\bm{x}) ~\text{s.t.}~ \bm{g}(\bm{x}) \leq \bm{B} \right\},
    \label{eqn:objective}
\end{equation}
where $\bm{x}$ is the design variable vector of $[0,1]^d$ described in Table \ref{tab:DV_CMDO}, $d$ the number of design variables, $\bm{g}$ is the inequality constraints described in Table \ref{tab:Cst_CMDO} and $\bm{B}$ the associated bounds.
Due to confidentiality reason $\bm{B}$ is not given.
Finally for the CMDO optimization problem (\ref{eqn:objective}), we consider here 12 design variables and 8 constraints.

Also note that the long-range cruise (i.e. nominal) mission range is satisfied implicitly by the workflow.
The baseline BRAC (see Figure \ref{fig:BRAC_3DView}) as defined in the CMDO workflow is unfeasible.
This is inconsequential, since that initial design is not used as a starting point by any of the optimizers.
Section~\ref{sec:opt} explains which starting point each optimizer uses instead.
For industrial reasons, the CMDO of BRAC cannot take more than $8$ hours.

\begin{table}[!ht]
	\centering
	\caption{Design optimization space definition.}
	\centerline{
	\begin{tabular}{l l l | l l l}
		\toprule
		\textbf{Variable} & \multirow{2}{*}{\textbf{Variable Name}} & \textbf{Variable} & \textbf{Variable} & \multirow{2}{*}{\textbf{Variable Name}} & \textbf{Variable} \\
		\textbf{index} & & \textbf{symbol} & \textbf{index} & & \textbf{symbol} \\
		\midrule
		$\bm{x}[0]$ & Engine scaling factor & DV\_0 & $\bm{x}[4-5]$ & Wing rear spar chord-wise locations & DV\_4-5\\
		$\bm{x}[1]$ & Wing aspect ratio & DV\_1 & $\bm{x}[6]$ & Wing sweep & DV\_6\\
		$\bm{x}[2]$ & Wing area & DV\_2  & $\bm{x}[7]$ & Wing taper & DV\_7 \\
		$\bm{x}[3]$ & Wing inboard trailing edge sweep & DV\_3 & $\bm{x}[8-11]$ & Wing maximum thickness-to-chord ratios & DV\_8-11 \\
		\bottomrule
	\end{tabular}}
	\label{tab:DV_CMDO}
\end{table}

\begin{table}[!ht]
	\centering
	\caption{Constraints definition.}
	\begin{tabular}{l l l | l l l}
		\toprule
		\textbf{Constraint} & \multirow{2}{*}{\textbf{Constraint Name}} & \textbf{Bounds} & \textbf{Constraint} & \multirow{2}{*}{\textbf{Constraint Name}} & \textbf{Bounds} \\
		\textbf{index} & & \textbf{index} & \textbf{index} & & \textbf{index} \\
		\midrule
		$\bm{g}[0]$ & BFL & $\bm{B}[0]$ & $\bm{g}[4]$ & Climb performance & $\bm{B}[4]$\\
		$\bm{g}[1]$ & ICA & $\bm{B}[1]$ & $\bm{g}[5]$ & High-speed mission range & $\bm{B}[5]$  \\
		$\bm{g}[2]$ & Vref & $\bm{B}[2]$ & $\bm{g}[6]$ & Landing gear spacing & $\bm{B}[6]$ \\
		$\bm{g}[3]$ & Excess fuel & $\bm{B}[3]$ & $\bm{g}[7]$ & Wingtip chord & $\bm{B}[7]$  \\
		\bottomrule
	\end{tabular}
	\label{tab:Cst_CMDO}
\end{table}

\subsection{BRAC PMDO problem}

As mentioned earlier, the PMDO portion of this study will consider the aero-structural workflow, shown on the left half of Figure~\ref{fig:PMDO_Framework}.
Similar to CMDO, the objective is to minimize MTOW, as in problem \eqref{eqn:objective}.
The wing and winglet design variables considered are listed in Table~\ref{tab:DV_PMDO}.
The wing area and engine size are kept constant at the PMDO level, and the wing area of the CMDO optimum is used as an input to the plan-form generator. Similarly, the spar locations are fixed using the CMDO result, since the high-lift systems integration is not considered.
The constraints imposed during PMDO are listed in Table~\ref{tab:Constraints_PMDO}.
The volume availability factor is a measure of the fuel volume available in the wing compared to the volume required to meet the range requirements.
The maximum wingtip twist deformation is a proxy for a dynamics constraint, and is enforced to prevent an overly flexible wing.
Again, for confidentiality reasons, the variable and constraint bounds are omitted.
Finally for the PMDO optimization problem (\ref{eqn:objective}), we consider here 19 design variables and 5 constraints which is representative of a typical PMDO industrial application.

\begin{table}[!ht]
	\centering
	\caption{PMDO design optimization space definition.}
	\centerline{
	\begin{tabular}{l l l | l l l}
		\toprule
		\textbf{Variable} & \multirow{2}{*}{\textbf{Variable Name}} & \textbf{Variable} & \textbf{Variable} & \multirow{2}{*}{\textbf{Variable Name}} & \textbf{Variable} \\
		\textbf{index} & & \textbf{symbol} & \textbf{index} & & \textbf{symbol} \\
		\midrule
		$\bm{x}[0]$ & Wing span & DV\_0 & $\bm{x}[8-13]$ & Wing max thickness-to-chord ratios & DV\_8-13 \\
		$\bm{x}[1]$ & Wing leading edge sweep & DV\_1 &  $\bm{x}[14]$ & Wing inboard trailing edge sweep & DV\_14 \\
		$\bm{x}[2]$ & Wing break loation & DV\_2 & $\bm{x}[15]$ & Winglet cant & DV\_15 \\
		$\bm{x}[3]$ & Wingtip chord & DV\_3 & $\bm{x}[16]$ & Winglet span & DV\_16  \\
		$\bm{x}[4-7]$ & Wing twist distribution & DV\_4-7 & $\bm{x}[17-18]$ & Winglet twist & DV\_17-18  \\	
		\bottomrule
	\end{tabular}}
	\label{tab:DV_PMDO}
\end{table}

\begin{table}[!ht]
	\centering
	\caption{PMDO constraints definition.}
	\begin{tabular}{l l l | l l l}
		\toprule
		\textbf{Constraint} & \multirow{2}{*}{\textbf{Constraint Name}} & \textbf{Bounds} & \textbf{Constraints} & \multirow{2}{*}{\textbf{Constraint Name}} & \textbf{Bounds} \\
		\textbf{index} & & \textbf{index} & \textbf{index} & & \textbf{index} \\
		\midrule
		$\bm{g}[0]$ & Volume availability factor & $\bm{B}[0]$ & $\bm{g}[3]$ & Landing gear spacing & $\bm{B}[3]$\\
		$\bm{g}[1]$ & $C_{L_{max}}$ retracted & $\bm{B}[1]$ & $\bm{g}[4]$ & Wingtip twist deformation & $\bm{B}[4]$  \\
		$\bm{g}[2]$ & Cruise deck angle & $\bm{B}[2]$ &  &  &  \\
		\bottomrule
	\end{tabular}
	\label{tab:Constraints_PMDO}
\end{table}

\section{Optimizers}
\label{sec:opt}

In this section, we introduce six optimizers to solve the following constrained problem:
\begin{equation}
    \min\limits_{\bm{x} \in \Omega} \left\{ f(\bm{x}) ~~\mbox{s.t.}~~ \bm{g}(\bm{x})\geq 0 \right\},
    \label{eq:opt_prob}
\end{equation}
where $f: \mathbb{R}^d \mapsto \mathbb{R}$ is the objective function, $\bm{g}:\mathbb{R}^d \mapsto \mathbb{R}^m$ gives the inequality constraints which are expensive to evaluate and $\bm{x} \in \Omega \subset \mathbb{R}^d $ is the vector of design variables.
Two of the optimizers are included in the Isight software, which is mandatory to use because of CMDO and PMDO frameworks. We developed an Isight interface for the SEGOMOE python tool-box in which the other optimizers are implemented. 

\subsection{Bayesian optimizers}
This section introduces the \textit{constrained Bayesian optimization} (CBO) framework \cite{MockusBayesianmethodsseeking1975,JonesEfficientglobaloptimization1998} that aims to solve the optimization problem~\eqref{eq:opt_prob} with a minimal number of calls.
To do so, one uses \textit{Gaussian Process} (GP) (also known as Kriging) \cite{RasmussenGaussianprocessesmachine2006,Krigestatisticalapproachbasic1951} trained with a pre-computed \textit{design of experiments} (DoE) (i.e. set of designs evaluated on the objective and constraints functions) of $l$ points.
GPs are then used to provide, with a cheap computational cost, a prediction $\mu_s^{(l)}: \mathbb{R}^d \mapsto \mathbb{R}$ and an associated uncertainty $\sigma_s^{(l)}: \mathbb{R}^d \mapsto \mathbb{R}$ for each point of $\bm{x} \in \Omega$ where $s : \mathbb{R}^d \mapsto \mathbb{R}$ can be either $f$ and $g_i$ for a given constraint component $i$. 
Concerning the objective function, these information are combined in an acquisition function  $\alpha_f^{(l)}: \mathbb{R}^d \mapsto \mathbb{R}$ \cite{frazier2018tutorial,Bartoliadaptivemodeling2019,WangMaxvalueentropysearch2017} coding the trade-off between exploration of the highly uncertain domain that can hide a minimum and exploitation of the minimum of the GP prediction. 
For the constraints, these information are joined to produce a feasibility criterion $\alpha_{\bm{g}}^{(l)}: \mathbb{R}^d \mapsto \mathbb{R}^m$ \cite{frazier2018tutorial,lam2015multifidelity,priem2019use} which is generally explicit.
The point $\bm{x}^{(l+1)}$, solving the constrained maximization trade-off sub-problem: 
\begin{equation}
    \bm{x}^{(l+1)}=\arg\min\limits_{\bm{x} \in \Omega} \left\{ \alpha_f^{(l)}(\bm{x}) ~~\mbox{s.t.}~~ \bm{x}\in \Omega_{\bm{g}}^{(l)} \right\},
    \label{eq:vst_prob}
\end{equation}
where $\Omega_{\bm{g}}^{(l)}$ is the approximated feasible domain defined by the feasibility criterion $\bm{\alpha}_{\bm{g}}^{(l)}$, is thus iteratively added to the DoE until a maximum number of iterations max\_nb\_it is reached.
The solution provided by CBO to the problem~\eqref{eq:opt_prob} is eventually the best point in the DoE (i.e with the minimal feasible value of $f$). 
The main steps of CBO are finally summarized in Algorithm~\ref{alg:BO}.
\begin{algorithm}[ht!]
     \begin{algorithmic}[1]
        \INPUT{Objective and constraints functions, initial DoE for objective and constraints, a maximum number of iterations max\_nb\_it\;}
        \FOR{$l = 1$ \TO \mbox{max\_nb\_it}}
            \STATE {Build the surrogate models using GPs\;}
            \STATE {Find $\bm{x}^{(l+1)}$ a solution of the enrichment maximization sub-problem\;}
            \STATE {Evaluate the objective and constraints functions at $\bm{x}^{(l+1)}$\;}
            \STATE {Update the DoE\;}
        \ENDFOR
        \OUTPUT{The best point found in the DoE\;}
    \end{algorithmic}
    \caption{The constrained Bayesian optimization framework.}
    \label{alg:BO}
\end{algorithm}

In this context, \citet{Bartoliadaptivemodeling2019} implemented a CBO python tool-box based on the \textit{super efficient global optimization} (SEGO) algorithm of \citet{SasenaExplorationmetamodelingsampling2002} which has been enhanced by the use of \textit{mixture of experts} (MOE) \cite{hastieelements2005, bettebghor2011surrogate}, the \textit{kriging with Partial Least Squares} (KPLS) for high dimensional problems~\cite{BouhlelImprovingkrigingsurrogates2016}, additional acquisition functions (e.g. WB2S \cite{Bartoliadaptivemodeling2019}) for highly multimodal objective functions, supplementary feasibility criteria (e.g. the \textit{upper trust bound} (UTB) \cite{lam2015multifidelity,priem2019use}) for non-linear constraint functions and a multiprocessing ability to speed-up the optimization process. 
The tool-box \cite{Bartoliadaptivemodeling2019} is thus named the \textit{super efficient global optimization with mixture of experts} (SEGOMOE) tool-box.
We also developed an interface of the python tool-box to allow its use in the Isight software in which BRAC is implemented. 

In this paper, the four following variants of SEGO (available with the SEGOMOE tool-box) are tested to show their common ability on industrial MDO problems:
\begin{itemize}
    \item SEGO \cite{SasenaExplorationmetamodelingsampling2002}: it can be seen as a basic implementation of the super efficient global optimization framework, it was developed for standard constrained optimization problems.
    \item SEGO-UTB \cite{priem2019use}: it includes a decreasing upper trust bound in the SEGO framework. In the context of this paper and to encourage the exploration of the feasible domain, we used an exponential decrease for the constraints learning rate, see \cite[Fig.  3]{priem2019use}.
    \item SEGOMOE \cite{Bartoliadaptivemodeling2019}: it combines the use of multiple models (local GPs) with the SEGO framework in order to mitigate  the high non-linearity of the targeted optimization problem. 
    \item SEGOMOE-UTB: similarly to SEGO-UTB, it combines the use of the UTB strategy and the SEGOMOE framework. Again, we choose to work with an exponential decrease for the constraints learning rate.
\end{itemize}
We note that for all the four variants we used the WB2S acquisition function combined with the KPLS models as the dimension of the design space is larger than 12. 

\subsection{Isight optimizers}

The Isight software \cite{vanisight2010} is delivered with many ready-to-use constrained optimizers.
In this paper, we focus on two of the popular derivative-free Isight constrained optimizers: the \textit{evolutionary optimization algorithm} (Evol) \cite{schwefelevolutionsstrategie1975} and a variant of the Pointerdog algorithm (Pointer-2)  \cite{vanisight2010}. 

Evol is an evolutionary strategy that mutates iteratively the best known point 
 by adding a normally distributed  perturbation to the design variables. 
The standard deviation of the normal distribution is adapted during the optimization process to solve the regarded problem
with a minimal number of evaluations. 
This simple algorithm is enhanced by different features such as: (1) a repeat calculation check to ensure that all the points computed are different, (2) a standard deviation expansion process when the same point is always evaluated, (3) a consecutive variable search allowing the exploration in a single canonical direction of the design space, and (4) a parallel execution accelerating the optimization process. 

The Pointer-2 strategy is an optimization framework based on an Isight proprietary algorithm that managed a set of well-known optimizers: Evol \cite{schwefelevolutionsstrategie1975}, the Hooke-Jeeves direct search method \cite{hookedirect1961}, the \textit{non-linear programming by quadratic Lagrangian} (NLPQL) algorithm \cite{schittkowskinlpql1986}, the downhill simplex algorithm \cite{neldersimplex1965}, the \textit{multi-function optimization system tool} (MOST) technique \cite{vanisight2010} and the \textit{multi-objective particle swarm optimization} (MOPSO) algorithm \cite{coellohandling2004}.
The choice of the set of optimizers used in the optimization process is made with respect to a classification system using the information available on the problem (i.e. often given by the user). 
The best optimizer, and its settings, are then updated thanks to the information collected all along the optimization process.
Furthermore, the optimization method can be used in two ways.
First, the optimizers and their settings are selected to have the higher improvement leading to the best solution in the shortest time. 
On the contrary, one searches a robust solution to uncertainties which leads to a slower optimization process due to a higher number of calls.
The Pointer-2 strategy periodically performs a surrogate based optimization to speed up the convergence.
In fact, the Pointer-2 is able to solve a wide range of constrained optimization problems and thus allows non-specialist to solve an optimization problem only providing the objective and constraints functions, the design variables, and a targeted optimization time.

\section{Results}

In this section, we introduce and comment the different tests realized on the BRAC CMDO and PMDO problems. 
We also recall the aim here is to find the best solution to the optimization problem, the fastest, and the different optimizers are tested in this scope.
In the following, the six previously introduced optimizers (see Section~\ref{sec:opt}) are tested on the BRAC CMDO and PMDO.
For the sake of comparison, we will produce convergence and parallel plots to assess the efficiency of the SEGOMOE tool-box solvers compared to the Isight optimizers. 

\subsection{BRAC CMDO results}

\subsubsection{Tests details}
\label{ssec:cmdo:details}

The optimizers, introduced in Section~\ref{sec:opt}, are compared with the following test plan. 
We perform $10$ independent runs for each solver using $10$ different initial DoEs build with the Latin hyper-cube sampling method. 
For each run, all the solvers are initiated with the same DoE. 
The size of the initial DoEs is set to $n_{start} = 13$ (i.e $d+1$ where $d$ is the dimension of BRAC CMDO problem).
Note that Evol and Pointer-2 need a single point to launch the optimization process, we thus provide the point with the best valid value of the DoE.
If there is no valid value in the DoE, the point with the minimal constraints violation is provided.

The maximum number of evaluations of the SEGO-like solvers is set to $\text{max\_it\_nb}=227$ (i.e. $20d - n_{start}$ meaning a total of $20d = 240$ evaluations is performed).
Concerning Evol and Pointer-2, we used the historical options settings of the CMDO framework.
The maximum number of evaluations of Evol is set to $\text{max\_it\_nb} = 960$.
For each run, Evol thus performed $973$ function evaluations, which takes approximately 8 hours to perform.
We respectively fix the maximum allowable job time and the topography type options of Pointer-2 to $8$ hours and \textit{nonlinear}. 
All the remaining options of Evol and Pointer-2 are kept as default.
Finally, the best solution among the ten optimizations performed for each optimizer is kept to initiate the BRAC PMDO problem as explained in Section~\ref{ssec:PMDO}.

\subsubsection{Convergence plots}
\label{sssec:cmdo:cv_plot}

To assess the performance of the introduced optimizers, we build two kinds of convergence plots. 
The first one, named \textit{evaluation convergence plot}, displays the average and the standard deviation of the best valid value over the $10$ runs for increasing number of evaluations. 
The best valid value is defined as the best valid value if there is, at least, one valid point in the DoE, otherwise, a penalization replaces the obtained invalid value.
The penalization is the highest feasible value of the objective function ever found. 
Because of confidentiality reasons, we scale the value of the convergence plots between $0$ and $1$.
The second convergence plot, named \textit{time convergence plot}, also shows the average and the standard deviation of the best valid value over the $10$ runs along the optimization time.
The penalization is kept the same as in the evaluation convergence plot. 

\begin{figure}[htp]
    \centering
    \subfloat[Evaluation convergence plot. \label{fig:CV_CMDO:Ite}]{\includegraphics[width=0.45\textwidth]{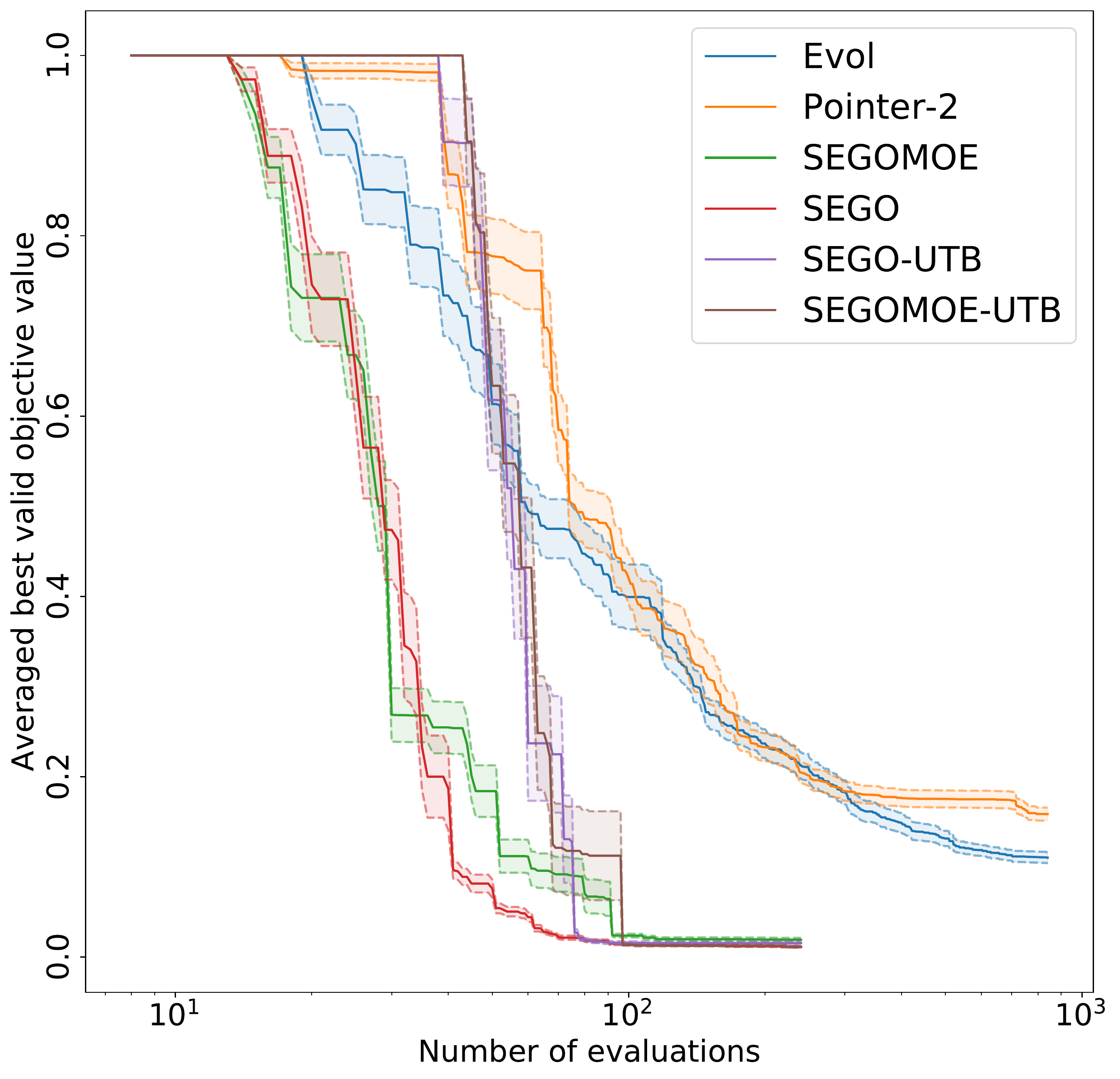}} 
    \subfloat[Time convergence plot. \label{fig:CMDO:time}]{\includegraphics[width=0.45\textwidth]{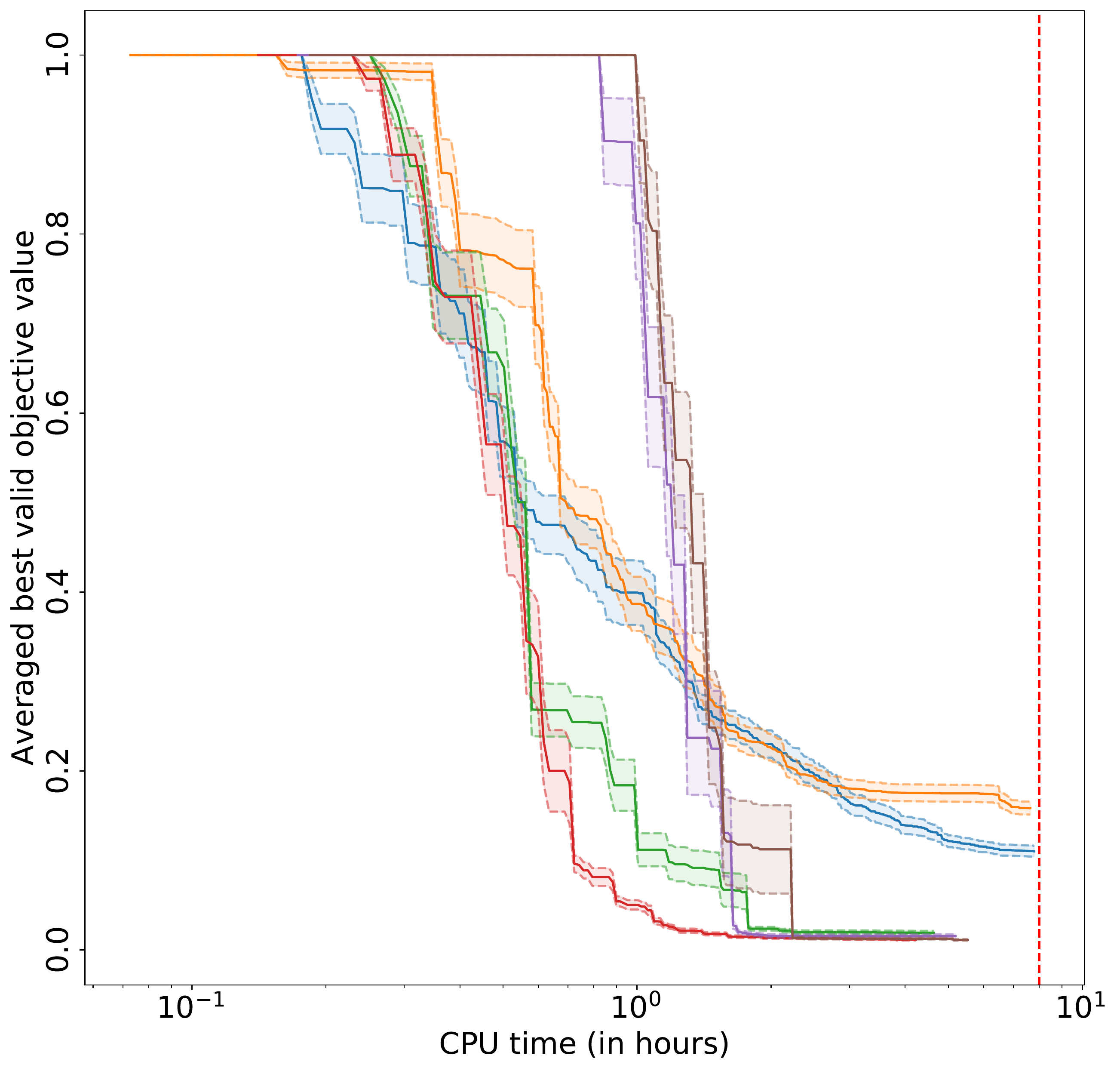}}
    \caption{Convergence plots for the CMDO of BRAC. In the time convergence plot, the vertical red-dashed line indicates the maximum CPU time allowed for an industrial optimization process (8 hours for this test case).}
    \label{fig:CV_CMDO}
\end{figure}

Figure~\ref{fig:CV_CMDO} shows that SEGO-like solvers clearly outperform Evol and Pointer-2 in term of the optimum value. 
In term of number of evaluations, Figure~\ref{fig:CV_CMDO:Ite} displays that SEGO is converging the fastest to the optimum value even if SEGOMOE has a fastest convergence rate at the beginning of the optimization. 
Furthermore, note that SEGO-UTB and SEGOMOE-UTB are not converging as fast as SEGO and SEGOMOE because of their extensive exploratory behaviour.
When looking to the optimization time, Figure~\ref{fig:CMDO:time} reveals that the SEGO-like solvers converge in approximately $2$ hours meanwhile Evol and Pointer-2 does not converge after $8$ hours. 

To conclude, these convergence plots have shown that the SEGO-like solvers provide a better solution in less running CPU-time than Evol and Pointer-2. 
We also note that the use of the UTB option within SEGO and SEGOMOE seems to not lead to any improvement on the obtained performance which suggests that the exploration of the feasible domain was not a difficult task to handle on this test case.
 
\subsubsection{Parallel plots}
\label{sssec:cmdo:para}

The parallel plots introduced in the following depict the behaviour of the tested solvers along the optimization process in which targeted values are displayed (e.g., the explored design variables).
Here, the plotted values are the number of iterations, the design variables values, the objective function value and the constraints violation. 
Furthermore, several colors are used to distinguish the reference design in red (i.e. the best feasible design found so far by all the tested optimizers, here SEGO), the optimum design found by the regarded solver in black, the feasible explored designs in green, and the unfeasible designs in blue.
Due to the stochastic nature of our tests, we build our parallel plots using a median run in the following way.
For each run of the optimizer, we store the best valid objective value; if none of the runs converges to a feasible point, we collect the minimal violation explored by the optimizer.
The median run is then selected based on the stored values for all runs.

\begin{figure}[p]
    \centering
    \subfloat[Evol. \label{fig:pp_cmdo:Evol}]{\includegraphics[height=0.26\textheight]{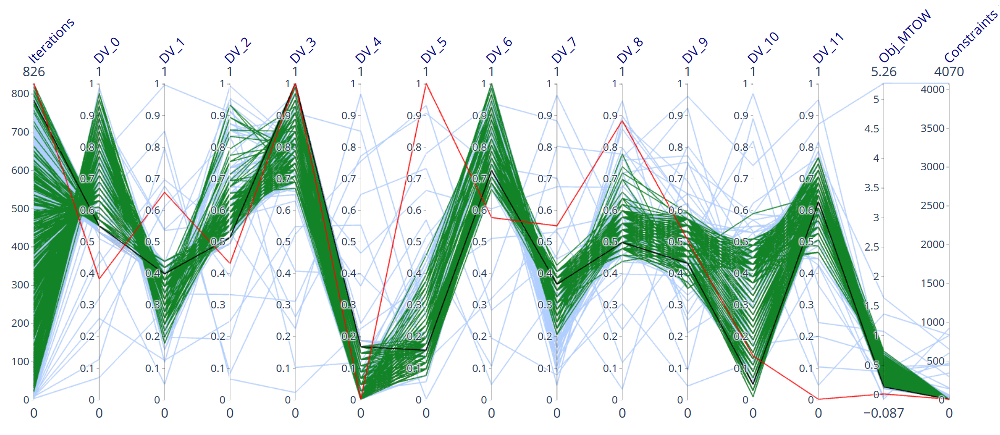}} \\
    \subfloat[Pointer-2. \label{fig:pp_cmdo:pointer}]{\includegraphics[height=0.26\textheight]{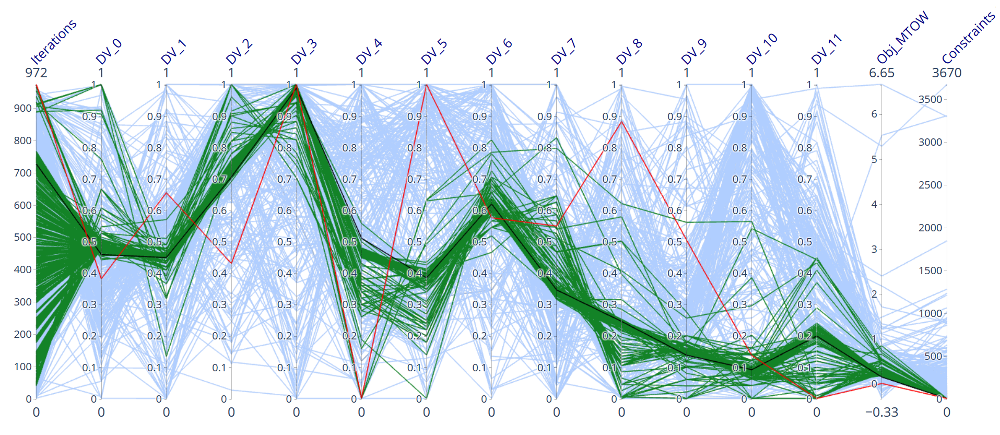}} \\
    \subfloat[SEGO. \label{fig:pp_cmdo:SEGO}]{\includegraphics[height=0.26\textheight]{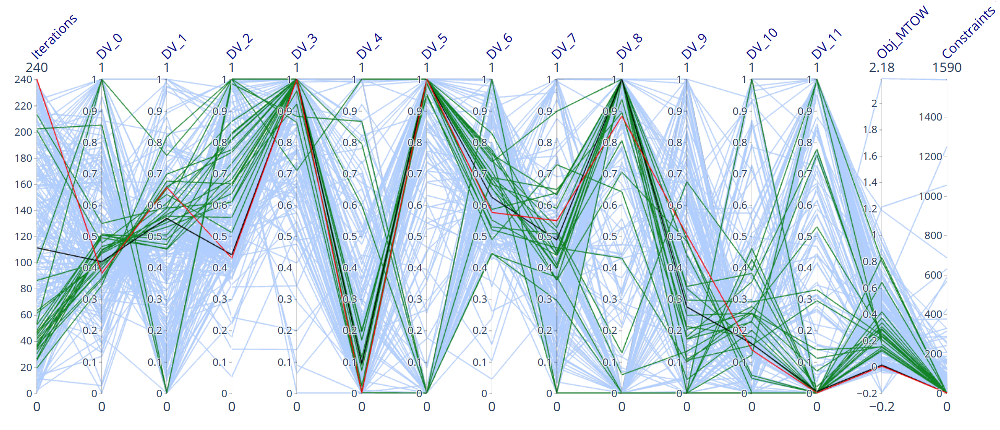}}
    \caption{Parallel plots using the median run for the CMDO of BRAC for Evol, Pointer-2 and SEGO. In blue: the unfeasible designs; in green: the feasible designs; in black: the optimum; in red: the reference design.}
    \label{fig:pp_cmdo}
\end{figure}

\begin{figure}[p]
    \centering
    \subfloat[SEGO-UTB. \label{fig:pp_cmdo:SEGO-UTB}]{\includegraphics[height=0.26\textheight]{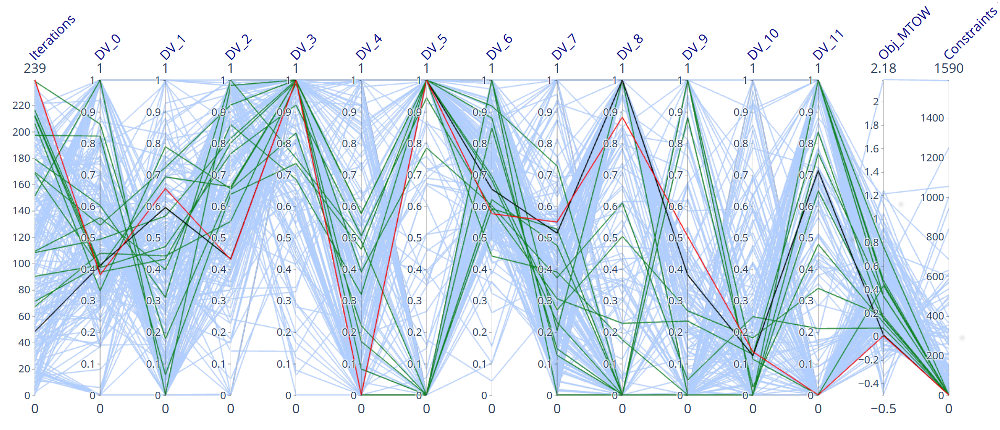}} \\
    \subfloat[SEGOMOE-UTB. \label{fig:pp_cmdo:SEGOMOE-UTB}]{\includegraphics[height=0.26\textheight]{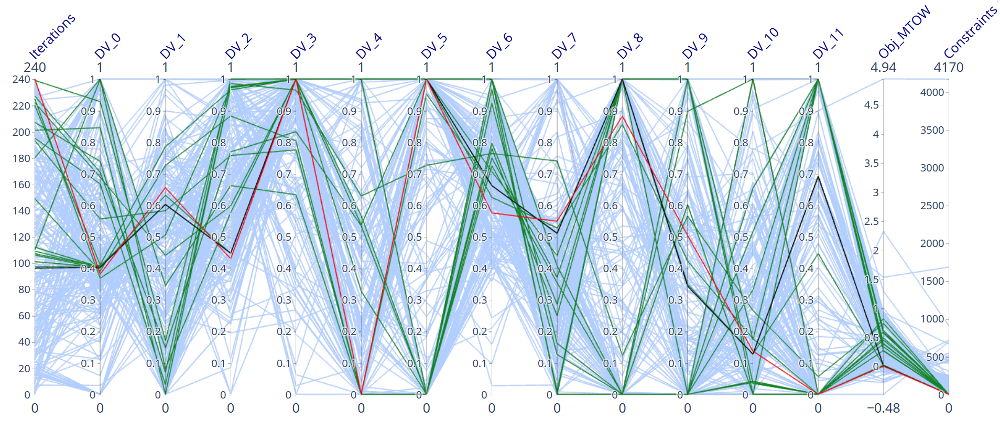}} \\
    \subfloat[SEGOMOE. \label{fig:pp_cmdo:SEGOMOE}]{\includegraphics[height=0.26\textheight]{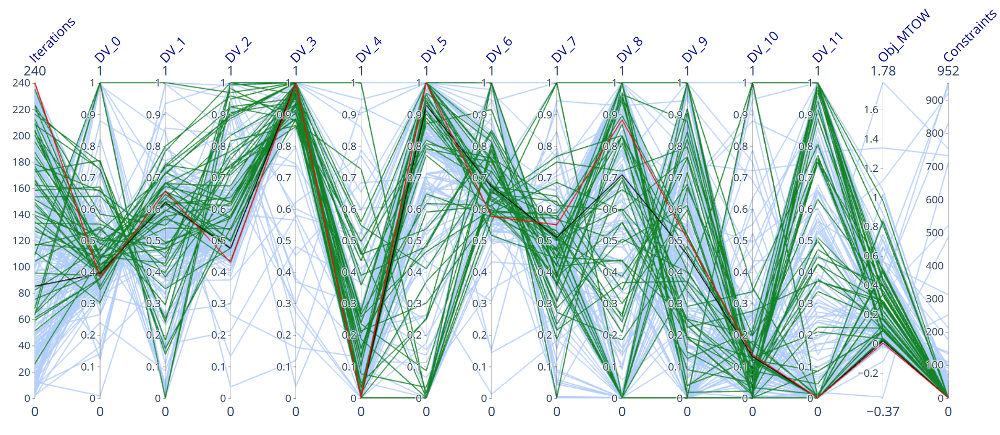}}
    \caption{Parallel plots using the median run for the CMDO of BRAC for SEGOMOE, SEGOMOE-UTB and SEGO-UTB. In blue: the unfeasible designs; in green: the feasible designs; in black: the optimum; in red: the reference design.}
    \label{fig:pp_cmdo_2}
\end{figure}

Figures~\ref{fig:pp_cmdo} and \ref{fig:pp_cmdo_2} display the parallel plots of the median run of the tested optimizers. 
First, note that all the SEGO-like optimizers are converging to the optimum. 
On the contrary, Evol and Pointer-2 do not converge to the optimum, as depicted by Figures~\ref{fig:pp_cmdo:Evol} and \ref{fig:pp_cmdo:pointer}. Furthermore, Evol is not able to explore the entire domain as implied by the grouped green lines of Figure~\ref{fig:pp_cmdo:Evol}.
Secondly, note the different solutions of the SEGO-like solvers even if they all converge approximately to the same objective function value.
Indeed, the values for DV\_8 and DV\_11 (i.e. wing maximum thickness-to-chord ratios) are different from an optimizer to another.
This can be due to either a local optimum or an inactive design variable (i.e. the objective function does not change along the design variable). 
From the expert point of view, the behavior observed on the BRAC CMDO problem is due to inactive design variables.

\subsection{PMDO results}

\subsubsection{Tests details}

The optimizers are again compared on the BRAC PMDO problem. 
Because of the computation time of the BRAC PMDO problem (ca. $25min$), we only performed one optimization for each solver. 
Each of the optimizers is initiated with the same DoE of $n_{start}=20$ sample points using the Latin hyper-cube sampling method.
To mimic the introduced MDO process, we also add to the initial DoE the point corresponding to the best solution $\bm{x}^*$ of all the BRAC CMDO process. 
As in Section~\ref{ssec:cmdo:details}, Evol and Pointer-2 only need a single point to launch the optimization, we thus provide the best solution $x^*$ of all the BRAC CMDO cases.

The maximum number of evaluations of the $4$ SEGO-like solvers is set to $\text{max\_it\_nb}=170$ (i.e. $10d - n_{start} - 1$ meaning a total of $10d = 190$ evaluations). 
The maximum number of evaluations of Evol and Pointer-2 is set to $510$ to follow historical values used in the PMDO framework, and the parallel batch size is set to $4$ for both.
In addition, a \textit{smooth} topography type is selected for Pointer-2.
All the other settings are kept to default values.

\subsubsection{Convergence plots}

\begin{figure}[htp]
    \centering
    \subfloat[Evaluation convergence plot. \label{fig:CV_PMDO:Ite}]{\includegraphics[width=0.45\textwidth]{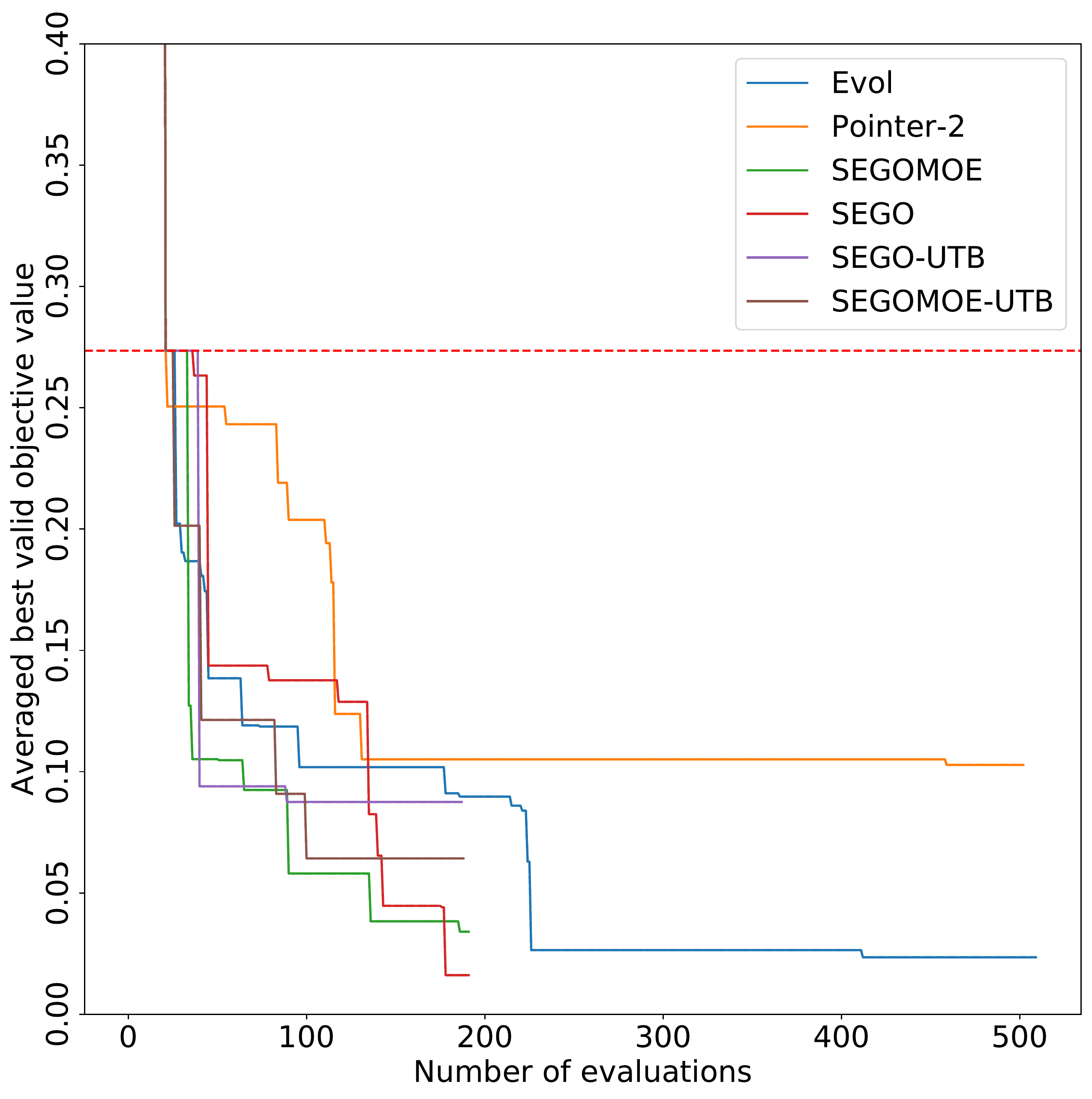}} 
    \subfloat[Time convergence plot. \label{fig:PMDO:time}]{\includegraphics[width=0.45\textwidth]{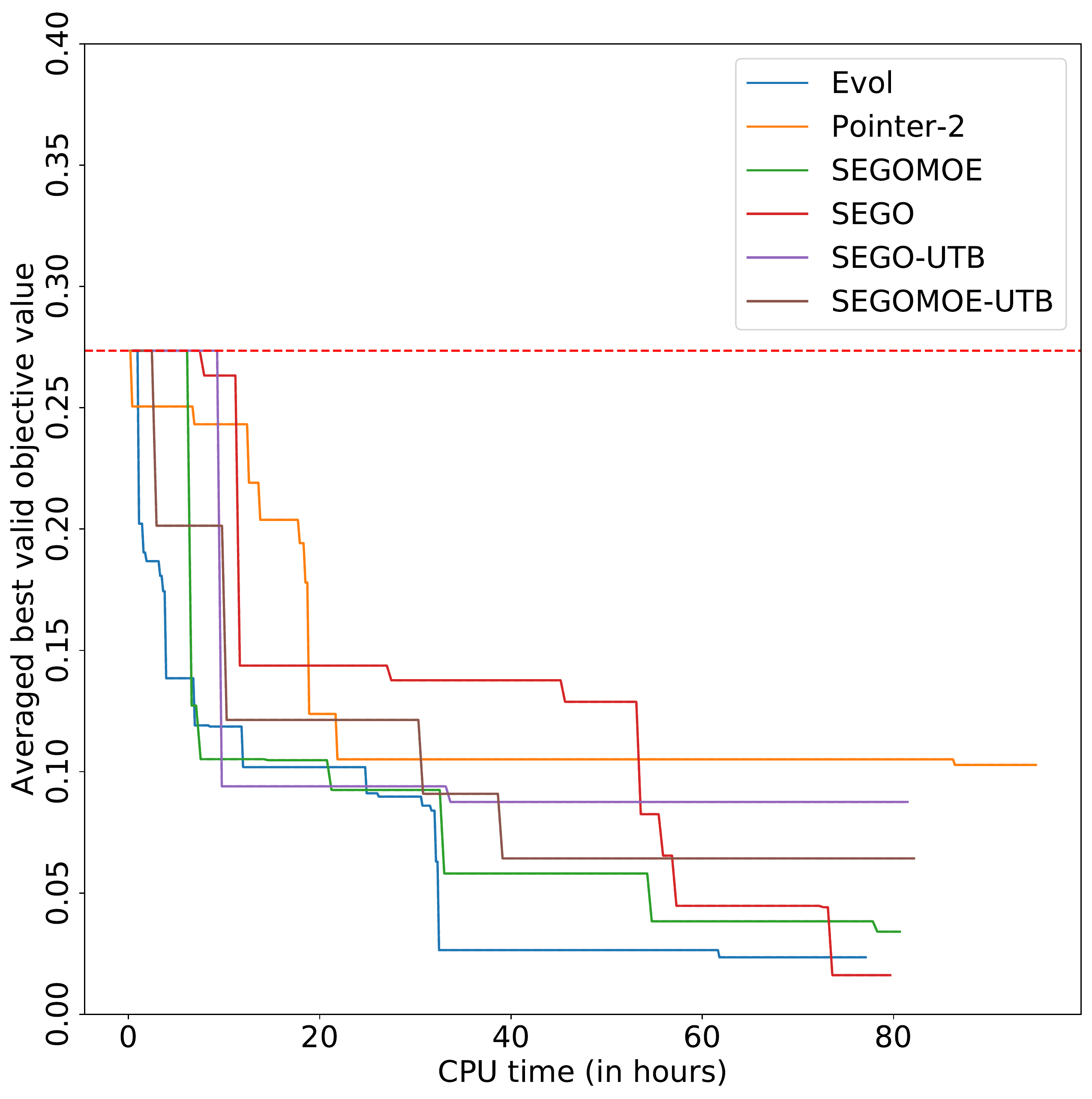}}
    \caption{Convergence plots for the PMDO of BRAC. The red dashed line is the value of the BRAC CMDO design in the BRAC PMDO process.}
    \label{fig:CV_PMDO}
\end{figure}

The convergence plots defined in this Section are slightly different from the ones introduced in Section~\ref{sssec:cmdo:cv_plot}.
Indeed, only one optimization is here performed for each of the $6$ optimizers. 
Thus Figure \ref{fig:CV_PMDO} displays the best valid values of the regarded solvers for increasing time or number of evaluations.  
First, Figure \ref{fig:CV_PMDO:Ite} shows that none of the solvers converge to the same optimum value, with the best value obtained with SEGO. 
Similar values are also achieved with Evol and SEGOMOE. 
Furthermore, Pointer-2 provides the worst value of all the solvers with an optimum at $0.1$. 
The other algorithms find better solution values than Pointer-2, but worse than SEGO, SEGOMOE and Evol.
Note that SEGO and SEGOMOE are able to find a similar solution as Evol with 30 fewer evaluations. 

Secondly, in terms of convergence time, Evol and Pointer-2 benefit from the batch evaluation of the objective function which is not the case for the SEGO-like solvers. 
Indeed, the 220 necessary evaluations of Evol to converge are done in less than 40 hours, whereas the SEGO-like solvers need 80 hours for 190 evaluations.

To conclude, Figure \ref{fig:CV_PMDO} shows that the UTB feasibility criterion is not useful for the BRAC CMDO problem, whereas SEGO and SEGOMOE show great performances with a small budget. 
Moreover, the batch evaluation capability of Evol speeds up the convergence time.

\subsubsection{Parallel plots}

\begin{figure}[p]
    \centering
    \subfloat[Evol. \label{fig:pp_pmdo:Evol}]{\includegraphics[height=0.26\textheight]{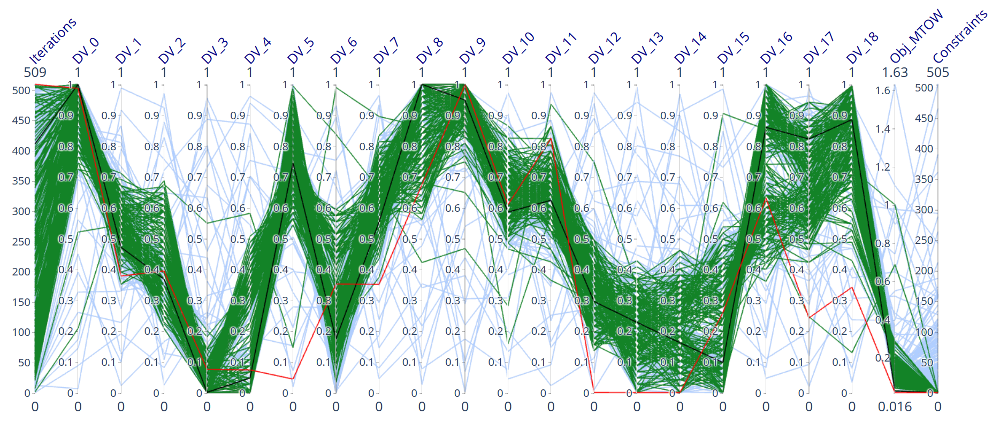}} \\
    \subfloat[Pointer-2. \label{fig:pp_pmdo:pointer}]{\includegraphics[height=0.26\textheight]{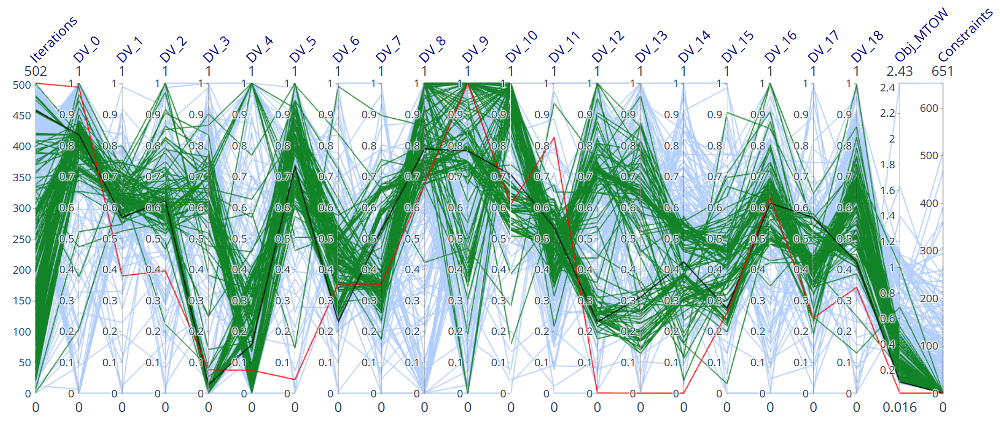}} \\
    \subfloat[SEGO. \label{fig:pp_pmdo:SEGO}]{\includegraphics[height=0.26\textheight]{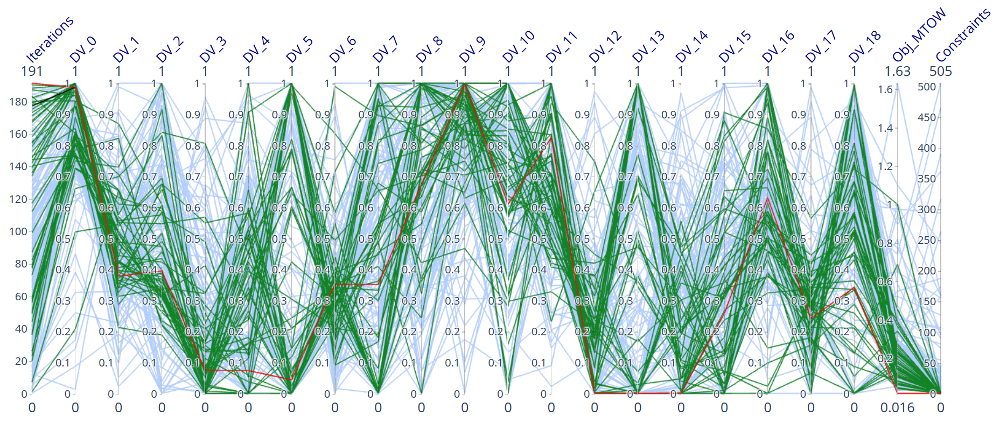}}
    \caption{Parallel plots for the PMDO of BRAC for Evol, Pointer-2 and SEGO. In blue: the unfeasible designs; in green: the feasible designs; in black: the optimum; in red: the reference design.}
    \label{fig:pp_pmdo}
\end{figure}

\begin{figure}[p]
    \centering
    \subfloat[SEGO-UTB. \label{fig:pp_pmdo:SEGO-UTB}]{\includegraphics[height=0.26\textheight]{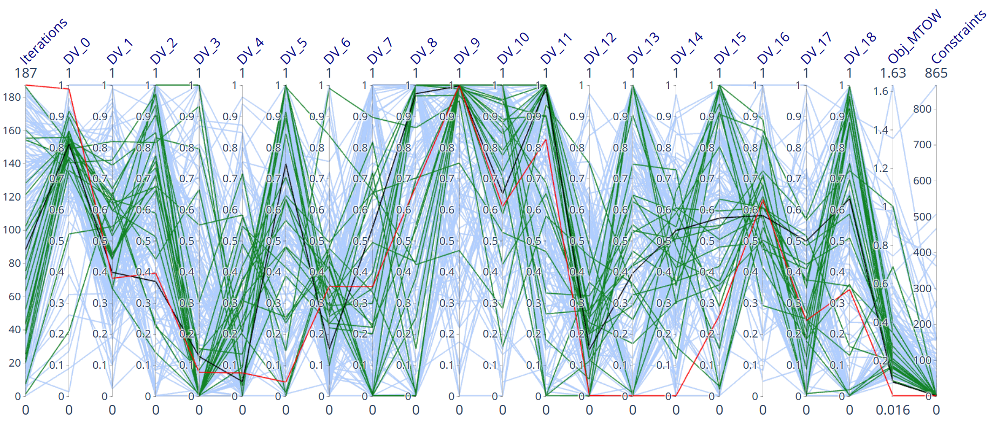}} \\
    \subfloat[SEGOMOE-UTB. \label{fig:pp_pmdo:SEGOMOE-UTB}]{\includegraphics[height=0.26\textheight]{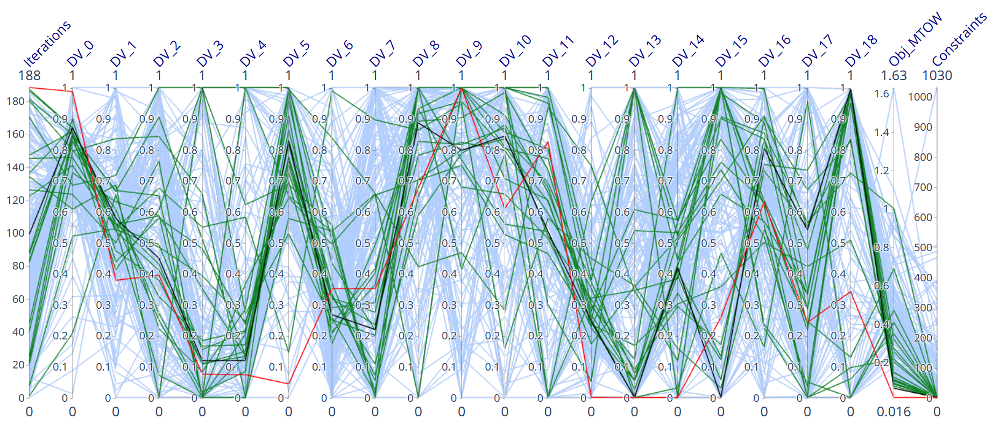}} \\
    \subfloat[SEGOMOE. \label{fig:pp_pmdo:SEGOMOE}]{\includegraphics[height=0.26\textheight]{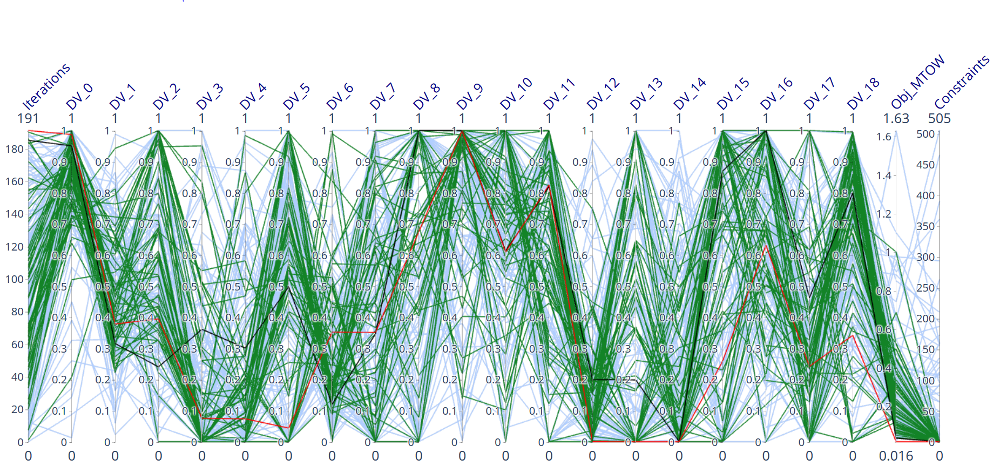}}
    \caption{Parallel plots for the PMDO of BRAC for SEGO-UTB, SEGO-UTB and SEGOMOE. In blue: the unfeasible designs; in green: the feasible designs; in black: the optimum; in red: the reference design.}
    \label{fig:pp_pmdo_2}
\end{figure}

Here, no run selections are needed for the parallel plots. Indeed, only a single run is performed for the BRAC PMDO problem for each solver.
They are produced in the same way as in Section \ref{sssec:cmdo:para}.
Figures \ref{fig:pp_pmdo} and \ref{fig:pp_pmdo_2} show the parallel plots for Evol, Pointer-2, SEGO, SEGOMOE, SEGO-UTB and SEGOMOE-UTB.

As mentioned, SEGO finds the best design, followed closely by Evol and SEGOMOE. These three solutions share similar main plan-form features, namely the wing span (DV\_0) and sweep (DV\_1), while there is some variation in the wing twist distribution (DV\_4-7) and maximum thickness-to-chord ratios (DV\_8-13), as shown in Figures~\ref{fig:pp_pmdo:Evol} and \ref{fig:pp_pmdo:SEGOMOE}. Only SEGO converges to the lower thickness bound for the two most outboard profiles (DV\_12 and DV\_13). 
The winglet has a secondary impact on the MTOW, and as a result even greater variation is seen in the optimized winglet parameters (DV\_15-18). 
The best solutions of the other three optimizers deviate even further from the SEGO optimum. It is possible that all the optimizers converge to different local optima, but the variability in the solutions is more likely due to incomplete convergence of the algorithms, since the number of iterations was limited to meet the time restrictions of an industrial application.

\section{Conclusion}

The multi-level optimization framework developed at Bombardier Aviation aims to help the design of efficient and competitive aircraft. 
The two optimization levels, implemented within the Isight software, were discussed in this paper: \textit{conceptual and preliminary multi-disciplinary optimization} (CMDO and PMDO).
The use of the ready-to-use Isight optimizers have shown great results in the past but within a minimum of 8 hours for the CMDO study case and a week for the PMDO one.
The SEGOMOE python tool-box was investigated in this paper and led to significant improvements on Bombardier research aircraft configuration test cases.
In particular, on the CMDO test case, we showed that all the tested solvers (implemented within the SEGOMOE tool-box) have outperformed two of the best Isight optimizers both in terms of the required computational effort as well as on the CPU running time. 
On the PMDO test case, the results are more contrasted. SEGO is able to provide a better solution than Evol in less evaluations but with more computational time. 
This is due to the batch evaluation capability of Evol. 
The batch evaluation capability of the SEGOMOE tool-box is a future improvement work to perform.

\section*{Acknowledgments}
This work is part of the activities of ONERA - ISAE - ENAC joint research group in a context of  a partnership between ONERA, ISAE-SUPAERO and Bombardier Aviation. In addition, we thank Jasveer Singh for his work on the BRAC CMDO specification.

\clearpage
\bibliography{Biblio}

\end{document}